\documentclass[12pt]{article}
\usepackage{graphicx}
\usepackage{float} 

\begin{document}
\begin{titlepage}

\title{Variations of the Energy of Free Particles in the pp-Wave
Spacetimes}

\author{J. W. Maluf$\,^{(1)}$, J. F. da Rocha-Neto$\,^{(2)}$, \\
S. C. Ulhoa$\,^{(3)}$, and F. L. Carneiro$\,^{(4)}$ \\
Instituto de F\'{\i}sica, 
Universidade de Bras\'{\i}lia\\
70.919-970 Bras\'{\i}lia DF, Brazil\\}
\maketitle
\bigskip
\bigskip

\begin{abstract}
We consider the action of exact plane gravitational waves, or pp-waves, on
free particles. The analysis is carried out by investigating the variations of 
the geodesic trajectories of the particles, before and after the passage of
the wave. The initial velocities of the particles are non-vanishing.
We evaluate numerically the kinetic energy per unit mass of the free particles,
and obtain interesting, quasi-periodic behaviour of the variations of the 
kinetic energy with respect to the width $\lambda$ of the gaussian that 
represents the wave. The variation of the
energy of the free particle is expected to be exactly minus the variation of 
the energy of the gravitational field, and therefore provides an estimation of
the local variation of the gravitational energy. The investigation is carried
out in the context of short bursts of gravitational waves, and of 
waves described by normalised gaussians, that yield impulsive waves in a 
certain limit. 
\end{abstract}
\thispagestyle{empty}
\vfill
\par

\bigskip
{\footnotesize
\noindent (1) wadih@unb.br, jwmaluf@gmail.com\par
\noindent (2) rocha@fis.unb.br\par
\noindent (3) sc.ulhoa@gmail.com\par
\noindent (4) fernandolessa45@gmail.com\par}

\end{titlepage}
\newpage

\section{Introduction}

Non-linear plane gravitational waves, or parallely propagated plane fronted
gravitational waves, are exact vacuum solutions of Einstein equations. They 
are also denoted simply by pp-waves. The space-time before and after the
passage of the wave is normally taken to be flat or nearly flat, for suitable
amplitudes of the waves. The metric tensor that describes the plane wave
space-time has a very simple structure. Assuming that the wave propagates along
the $z$ direction, the line element reads 

\begin{equation}
ds^2=dx^2+dy^2+2du\,dv+H(x,y,u)du^2\,,
\label{1}
\end{equation}
in $(u,v,x,y)$ coordinates. In the flat space-time region, we may identify

\begin{equation}
u = {1\over {\sqrt{2}}}(z-t)\,,
\label{2}
\end{equation}

\begin{equation}
v = {1\over {\sqrt{2}}}(z+t)\,.
\label{3}
\end{equation}

The coordinates $(x,y)$ parametrize the planes parallel to the wave front. The
line element depends essentially on a function $H(x,y,u)$.

The class of exact solutions that describe these waves is known as the Kundt 
family of space-times, and they were studied in great detail by Ehlers and Kundt
\cite{Ehlers,Ehlers-2} (see also Refs. \cite{Kramer,GP}). 
The dependence of the function $H$ on the 
variable $u$ is, to a large extent, arbitrary, a feature that is typical of 
solutions of wave equations. In view of this arbitrariness, the function $H$ 
may be modelled so that it represents a short burst of gravitational wave, in 
which case the $u$ dependence of $H$ may be given by smooth gaussians or 
derivatives of gaussians. The function $H$ must only satisfy \cite{Ehlers} 

\begin{equation}
\nabla^2 H=\biggl( {{\partial^2} \over{\partial x^2}}+ 
{{\partial^2} \over{\partial y^2}} \biggr)H=0\,.
\label{4}
\end{equation}
We assume that, far from the source, we may approximate a realistic 
gravitational wave by an exact plane wave, exactly as we do in the study of 
electromagnetic waves. The non-linear gravitational wave space-time is 
geodesicaly complete. The form above of the line element (\ref{1}), in the 
coordinates $(u,v,x,y)$, was first presented by  Brinkmann \cite{Brinkmann}.

The physical viability of these exact wave solutions has been supported by 
Penrose \cite{Penrose}, who argued that the waves are physically acceptable, in
similarity to the classical, source free, monochromatic plane 
electromagnetic waves. Penrose also addressed technical issues regarding the
establishment of a global Cauchy hypersurface in these space-times. According to
him, a light cone in an event $Q$ in the past (after the passage) of the 
wave, can never meet an event $R$ ahead (before the passage) of the 
wave. Thus, the Cauchy data on a hypersurface that contains the event $Q$ 
cannot give information for specifying the wave amplitude on $R$. But as we 
discussed above, the wave amplitude (contained in the function $H$) is 
necessarily an arbitrary function of $u$. Electromagnetic and gravitational 
waves are both assumed to travel at the same speed, the speed of light. The 
pp-waves do not violate any physical principle or property, such as causality,
for instance.

Non-linear gravitational waves have been recently studied in the analysis of 
the memory effect \cite{ZDGH1,ZDGH2,ZDH,ZDGH3,ZEGH}. 
The latter effect is established by the permanent displacement of free 
particles, after the passage of a wave. The displacement in the 
three-dimensional space may also be followed by a
variation of the velocities of the particles. The effect has been formulated 
by Zel´dovich and Polnarev  \cite{ZP}, and by Braginsky and Grishchuk \cite{BG},
even though Ehlers and Kundt \cite{Ehlers} evaluated the variation of the 
velocities of free particles after the passage of a pp-wave earlier. The memory
effect is relevant and conceptually important to future observations of 
gravitational waves in space, and to a better understanding of gravitational
waves in general.

The $u$ dependence of $H$ may be given by gaussians, as mentioned above, or
derivatives of gaussians. Normalised gaussians, such as 

\begin{equation}
{1\over {\lambda \sqrt{2\pi}}}\int_{-\infty}^{+\infty} du \;
e^{-{(u^2/ 2\lambda^2})} =1\;,
\label{5}
\end{equation}
lead to impulsive gravitational waves in the limit $\lambda \rightarrow 0$. The
gaussian becomes a delta function and the analysis of the geodesics
may be carried out analytically. By working with the limit 
$\lambda \rightarrow 0$ of the gaussian function, and not strictly with the 
delta functions themselves, one may overcome difficulties that arise from the 
ill defined products of delta functions. This is the procedure adopted in 
Refs. \cite{Steinbauer,Podolsky-2}, and which allows solving analytically the 
geodesic equations (see also Ref.\cite{ZDH}). 
We mention that the completeness of general pp-wave 
space-times has recently been considered in Ref. \cite{Saemann}, and pp-waves
with gyratons were investigated in Ref. \cite{Podolsky-3}.
Gyratonic pp-waves are not vacuum solutions of the field equations.
They are generated by a source that travels at the speed of light, and that is
described by the radiation density and by a quantity that represents the 
spinning character of the source. These waves yield the source free pp-waves
in a certain limit.

The $u$ dependence of $H$ in terms of derivatives of gaussians has been 
considered in the context of the memory effect, as for instance in Refs.
\cite{ZDGH1,ZDGH2}. The advantage of working specifically 
with the second derivative of the gaussian is that one does not need a 
multiplicative dimensional constant in $H$.
Recall that in Eq. (\ref{3}), $H$ must have dimension (length)$^{-2}$,
and therefore the second derivative of the gaussian, $(d^2/du^2)$ eliminates the
necessity of a dimensional amplitude. 

An interesting feature of non-linear plane gravitational waves is that their 
action on free particles may reduce the kinetic energy \cite{MRUC1}, as well as
the angular momentum of the particles \cite{MRUC2}. One would expect, 
{\it a priori}, that the energy of a particle would always increase after the
passage of the wave. This is obviously the case if the particles are initially
at rest. The increasing or decreasing
of kinetic energy of the free particles depends essentially on the initial 
conditions of the latter. Therefore, a pp-wave may remove energy from a
medium, and this feature may explain why a gravitational wave travels in space
for periods of time such as billions of years. If the gravitational wave were to
only transfer energy to the medium, which may be characterised as a cloud of
particles (an idea first proposed by Bondi \cite{Bondi}), then it would  be
difficult to understand how can a gravitational wave travel in space without
dissipating  for such long periods of time.

The feature mentioned above has
implications to the concept of gravitational energy-momentum. The
exchange of gravitational energy-momentum and angular momentum between the 
particle and the field takes place locally, i.e., where the free particles are
located. Therefore, it seems reasonable to assume that the increasing or
decreasing of gravitational energy also takes place locally. Since the kinetic
energy of the particles (before, during or after the passage of the wave, in a 
flat space-time region) cannot be removed by a coordinate transformation, the 
gravitational energy also cannot be removed by such a transformation (the 
removal of the gravitational field and, consequently, of the gravitational 
energy, by a coordinate transformation is a standard assumption against
the localization of the gravitational energy).

In this article we continue our analysis developed in Refs. \cite{MRUC1,MRUC2},
and search for discrete or periodic structures in the behaviour of a free
particle, in the pp-wave space-times. The particles in consideration are 
free, except that they are subject only to the gravitational field of the wave,
and the space-time is essentially flat outside the wave front (before and after
the passage of the wave), in view of the 
fall off of the gaussians. Our idea is that instead of looking for 
particular features of the gravitational field or gravitational 
energy, we investigate the properties of the free particles, and infer the 
corresponding behaviour (variation) of the gravitational field properties.
We find that the quantity $\Delta K \,$ = (final kinetic energy per unit mass) 
- (initial kinetic energy per unit mass) oscillates when varied with respect 
to $\lambda$, which is a quantity that is proportional to the width of the 
gaussian, as in Eq. (\ref{5}). The zeros of $\Delta K \,$ arise in discrete
values. In addition, we find that although the energy 
of a single particle changes continuously during the passage of the wave, 
there arise certain jumps and peaks in the expression of the kinetic
energy as a function of $u$. The increasing or decreasing of the kinetic energy
takes place by means of a sequence of peaks, that suggest the exchange of 
small packages of energy in a hypothetical quantum formulation of the problem.
We comment on and speculate about these peaks in the last sections of this 
article.

In section 2 we present the underlying mathematical formulation and the basic
geodesic equations for a particle in the pp-wave space-time. In section 3 we 
present the several relevant figures that we obtain by using the program 
MATHEMATICA. It seems that there does not exist exact, general 
solutions of the geodesic equations if the amplitude of the wave is given by
gaussians or derivatives of gaussians. For this reason, we resort to computer
programs to obtain graphic results. And, finally, in section 4 we discuss on the
results obtained in section 3, and speculate about a possible way to evaluate
the quantum of gravitational energy.

\section{The geodesic equations}

A non-linear plane gravitational wave may be written in several different forms
\cite{Kramer}. One possible form is the plane-fronted gravitational wave that 
we assume to propagate in the  $z$ direction, and which is given by 
Eq. (\ref{1}). This form for the line element was
first proposed by Brinkmann \cite{Brinkmann}. With the help of Eqs. 
(\ref{2}) and (\ref{3}), we rewrite the line element (\ref{1}) in $ (t,x,y,z)$ 
coordinates. It reads

\begin{equation}
ds^2=\biggl({H\over 2} -1\biggr)dt^2+dx^2+dy^2+
\biggl({H\over 2}+1\biggr) dz^2-H\,dt dz\,.
\label{6}
\end{equation}
We are assuming $c=1$.
The function $H$ must satisfy only Eq. (\ref{4}). As we mentioned earlier, the
dependence of $H$ on the retarded time $(-u)$ is arbitrary. The geodesic equations
in terms of the $t, x, y, z$ coordinates are \cite{JF} 

\begin{equation}
2\ddot{t} + \sqrt{2}H\ddot{u} + \sqrt{2}\dot{H}\dot{u} 
- {1\over \sqrt{2}}{\partial H\over \partial u}\dot{u}^2 = 0,
\label{7}
\end{equation}
\begin{equation}
2\ddot{x} - {\partial H\over \partial x}\dot{u}^{2} = 0,
\label{8}
\end{equation}
\begin{equation}
2\ddot{y} - {\partial H\over \partial y}\dot{u}^{2} =  0,
\label{9}
\end{equation}
\begin{equation}
2\ddot{z} +  \sqrt{2}H\ddot{u} + \sqrt{2}\dot{H}\dot{u} 
- {1\over \sqrt{2}}{\partial H\over \partial u}\dot{u}^2 = 0\,,
\label{10}
\end{equation}
where the dot represents derivative with respect to $s$.
From the first and fourth equations above we get

\begin{equation}
\ddot{z} - \ddot{t} = 0 \to \ddot{u} = 0 \to \dot{u}=
{1\over \sqrt{2}}(\dot{z} - \dot{t}) = constant.
\label{11}
\end{equation}
We assume the constant on the right hand side of the equation above to be $1$,
and thus consider $u$ to parametrize the geodesic curves.

Except for constant multiplicative factors, the function $H$ may be given by
the $+$ or $\times$ polarizations, according to

\begin{equation}
H=A_{+}(u)(x^2-y^2)\,,
\label{12}
\end{equation}

\begin{equation}
H=A_{\times}(u)\,xy\,,
\label{13}
\end{equation}
or by a linear combination of these quantities. 
The two expressions above of $H$ satisfy Eq. (\ref{4}). Both expressions lead 
to the same qualitative behaviour for the geodesics and kinetic energy of the
particles. In the present analysis we will consider the amplitudes $A_{+}(u)$ 
and $A_{\times}(u)$ to be given by regular gaussians, that represent short
bursts of gravitational waves, and by normalised gaussians (normalised not
necessarily to 1. We will choose multiplicative constants that yield 
satisfactory figures). The latter is typically given by 

\begin{equation}
A_{(+,\times)} \simeq
{1\over \lambda }\; e^{-{(u^2/ 2\lambda^2})}\,.
\label{14}
\end{equation}

The effect of a gravitational wave on a free particle is assumed to be very 
weak. The kinetic energy per unit mass $K$ of the free particles is considered
as

\begin{equation}
2K={1\over 2}(V_x^2 + V_y^2+ V_z^2)\;,
\label{15}
\end{equation}
where the velocities are
$V_x=dx/du$, $V_y=dy/du$ and $V_z=dz/du$. Thus, in view of Eq. (\ref{2}) we have

\begin{equation}
K={1\over 2}\biggl[ \biggl({{dx}\over {dt}}\biggr)^2+ 
\biggl({{dy}\over {dt}}\biggr)^2+\biggl({{dz}\over {dt}}\biggr)^2 \biggr]\,,
\label{16}
\end{equation}
which is, certainly, a valid expression before and after the passage of the 
wave, in which case the space-time is flat, or nearly flat.

\section{Kinetic energy and geodesics of free particles}

The particles that we consider in the present analysis are idealised 
entities that do not affect the gravitational field of the wave. Ehlers and 
Kundt \cite{Ehlers} already pointed out that, in principle, a more realistic 
approach would require to take into account the interaction between an ordinary
massive particle and the wave. For the time 
being, this is a very complicated issue to be addressed analytically, and even
numerically. But in view of the arbitrariness in the $u$ dependence of $H$, the
actual interaction between a massive particle and the wave will not change the
qualitative results obtained by just investigating ordinary geodesics in
space-time. We can formally make the gravitational field of the wave more 
intense than the field of a massive test particle, even though the gravitational
wave is taken to be weak.

In our previous investigations on this issue \cite{MRUC1,MRUC2}, we found that
the initial conditions are of extreme importance for obtaining certain 
variations, such as the increasing or decreasing of the kinetic energy of the
particles. So far, it is not clear what is the relation between the variations
of the geodesics, energy, momentum and angular momentum of the 
particles, on one hand, and the initial conditions, on the other hand. We 
will assume that the initial conditions for an astrophysical system like a star,
for instance, specially regarding the initial velocities, are roughly the same 
for all material components of the system (considering, of course, that the 
translational velocity of the star is much greater that the relative velocities
between its molecules).

According to Eq. (\ref{2}), a positive variation of $t$ corresponds to a 
negative variation of $u$. Thus, the increase of the time parameter $t$, for an
observer in the flat space-time before or after the passage of the wave, 
corresponds to the decreasing values of $u$. We will define the variation
$\Delta K$ of the kinetic energy per unit mass of the free particles by

\begin{equation}
\Delta K= K_f-K_i\,,
\label{17}
\end{equation}
where $K_f$ and $K_i$ are evaluated at $u=-200$ and $u=+200$, respectively, in
natural units. We also define the normalised variation $\Delta K_N $ according to

\begin{equation}
\Delta K_N= {{K_f-K_i}\over {K_f+K_i}}\,.
\label{18}
\end{equation}
where again $K_f$ and $K_i$ are evaluated at $u=-200$ and $u=+200$, 
respectively.

In the analysis below we will obtain the kinetic energy of the free particles,
and certain geodesics, by means of the program MATHEMATICA. As to our knowledge,
the general, exact solutions of the geodesic equations are not known, if the
amplitude of the wave is given by gaussians.
As we mentioned above, the geodesic equations and the kinetic energy of the 
free particles are highly dependent on the initial conditions. We will
consider 3 sets of initial conditions, all at $u=0$. These conditions
yield suitable graphic results. In natural units, they read

\begin{equation}
\textsl{I}:\ \ \ 
\rho_0=0.6\,, \ \ \ \phi_0=0\,, \ \ \ z_0=0\,, \ \ \ \dot{\rho}_0=0.2\,, \ \ \
\dot{\phi}_0=0\,, \ \ \ \dot{z}_0=0\,,
\label{19}
\end{equation}

\begin{equation}
\textsl{II}:\ \ \
\rho_0=0.6\,, \ \ \ \phi_0=0\,, \ \ \ z_0=0\,, \ \ \ \dot{\rho}_0=-0.2\,, \ \ \
\dot{\phi}_0=0\,, \ \ \ \dot{z}_0=0\,,
\label{20}
\end{equation}

\begin{equation}
\textsl{III}:\ \ \
\rho_0=10^{-5}\,, \ \ \ \phi_0=0\,, \ \ \ z_0=0\,, \ \ \ 
\dot{\rho}_0=10^{-5}\,, \ \ \   \dot{\phi}_0=0\,, \ \ \ \dot{z}_0=0\,.
\label{21}
\end{equation}

The $(\rho,\phi)$ variables are related to the $(x,y)$ variables by means of the
standard relations $x=\rho \cos\phi\,, \ \ y=\rho\sin\phi$. Thus, 

$$x^2-y^2=\rho^2(2\cos^2\phi -1)\,, \ \ \ xy={1\over 2}\rho^2 \sin(2\phi)\,.$$

\subsection{Non-normalised gaussians}

\begin{figure}[H]
	\centering
		\includegraphics[width=0.80\textwidth]{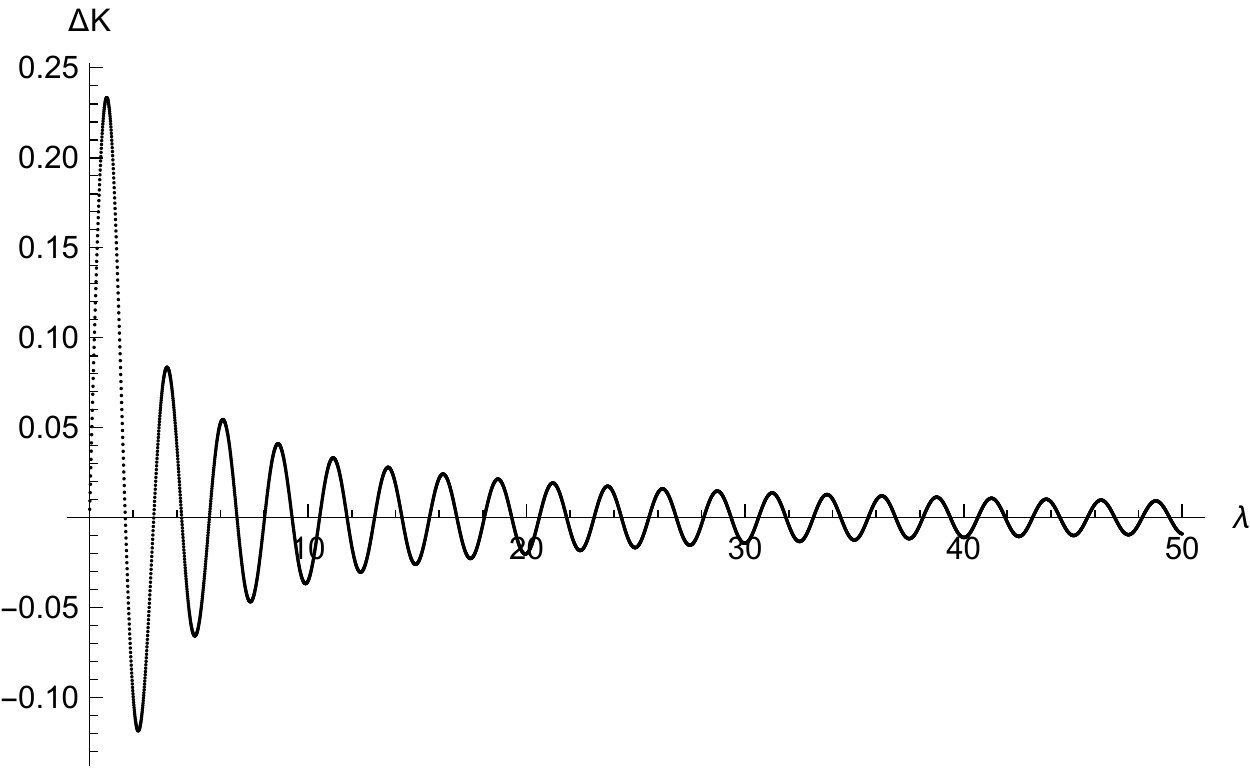}
	\caption{$\Delta K$ as a function of $\lambda$, considering $H_1$ and 
	initial 	conditions I.}
	\label{Figure-1}
\end{figure}

We will first address non-normalised gaussians, and choose two expressions for 
the function $H$. The expressions are 

\begin{eqnarray}
H_1&=&- \rho^2(u)\lbrack 2\cos^2(\phi(u))-1\rbrack \,e^{-u^2/\lambda^2}\,,
\label{22} \\
H_2&=&{1\over 2} \rho^2(u) \sin(2\phi(u))\,e^{-u^2/\lambda^2}\,.\label{23}
\end{eqnarray}
The quantity $\lambda$ is related to the width of the wave.
Differently from our previous analyses \cite{MRUC1}, we are now choosing
a minus sign in Eq. (\ref{22}), which just amounts to choosing $y^2-x^2$ instead
of $x^2-y^2$. We will plot the variation $\Delta K$ with respect to $\lambda$.
The result is presented in Figure (\ref{Figure-1}).

\begin{figure}[H]
	\centering
		\includegraphics[width=0.80\textwidth]{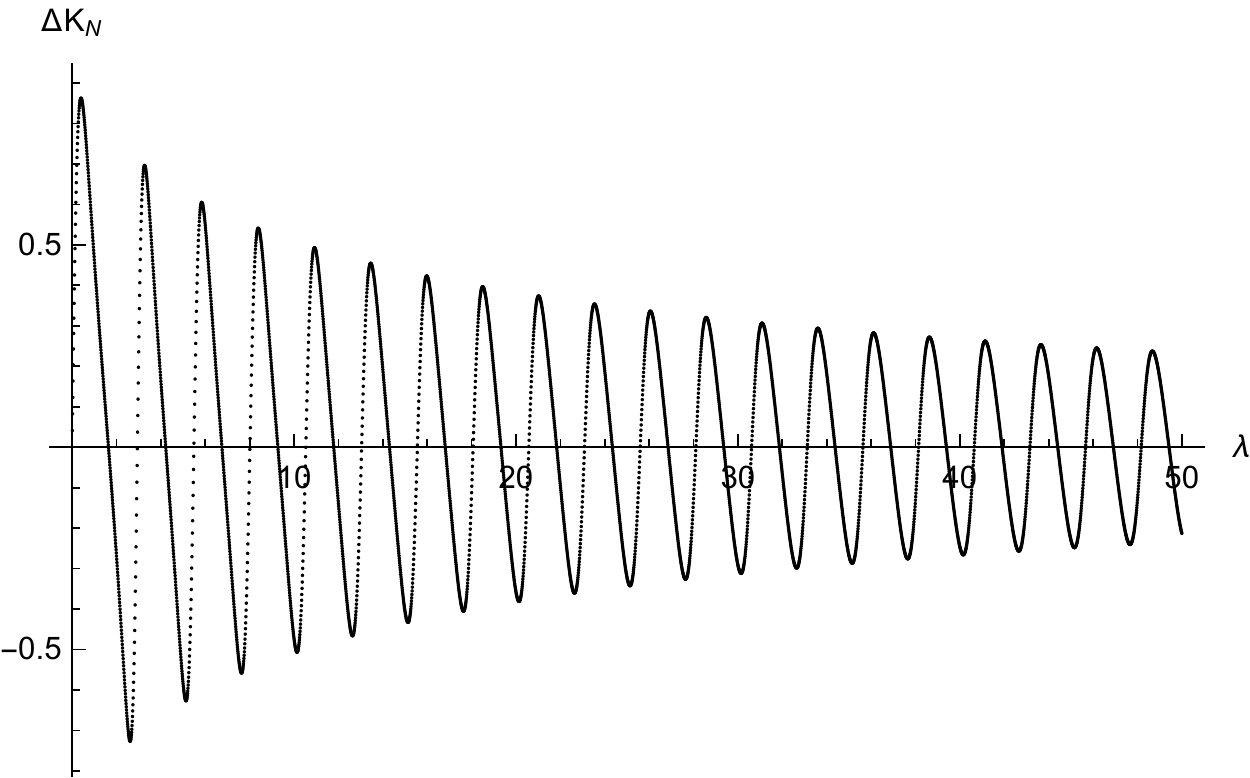}
	\caption{$\Delta K_N$ as a function of $\lambda$, considering $H_1$ and 
		initial conditions I.}
	\label{Figure-2}
\end{figure}

In natural units, $\lambda$ has dimension of length, and $\Delta K$ 
is dimensionless, since it is per unit mass.
Not only the initial conditions, also the value of the quantity $\lambda$ are
relevant for obtaining a decrease or increase of the kinetic energy of the 
particle. But in this case, we see that certain discrete values of $\lambda$ 
yield $\Delta K =0$. The quasi-periodic behaviour of the variation of the 
kinetic energy is more strengthened if we consider $\Delta K_N$ with respect
to $\lambda$. The result is very much similar, and is displayed in Figure 
(\ref{Figure-2}).

We see that a quasi-periodic structure is hidden in the field quantities and 
geodesic equations. Note that there are no periodic functions in Eqs. 
(\ref{7})-(\ref{10}). We also see that for certain small values of $\lambda$,
the impact on the particle is higher than for large values of $\lambda$.
This is the case when the ``sandwich'' wave is
concentrated in a small interval $\delta u$ around $u=0$. The figure above for 
$\Delta K_N$ may be given analytically, to a very good approximation, by the 
expression

\begin{equation}
\Delta K_N=f\,e^{-(a\lambda)/100}\sin \biggl( {{b\lambda}\over 100} +c\biggr)\,,
\label{24}
\end{equation}
where $f=0.55108$, $a=0.0201072$, $b=2.50067$ and $c=1.32005$, in natural units.
The maximal energy transfer between the particle and the field occur at the 
maxima and minima of $\Delta K_N$, and the corresponding values of $\lambda$ are 
given by

$${{b\lambda}\over 100}+c ={\pi \over 2} + n\pi\;,$$
where $n$ is an integer.

\begin{figure}[H]
	\centering
		\includegraphics[width=0.60\textwidth]{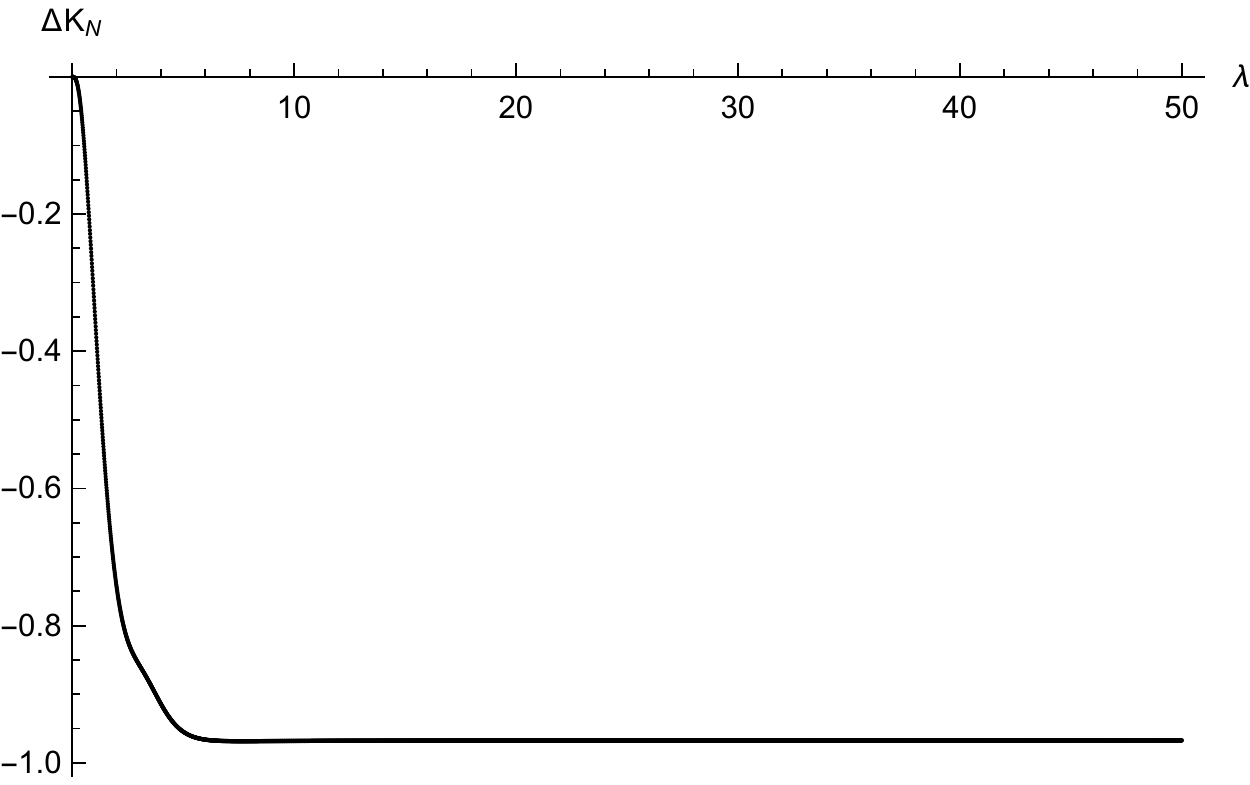}
	\caption{$\Delta K_N$ as a function of $\lambda$, considering $H_2$ and 
	     	initial conditions I. }
	\label{Figure-3}
\end{figure}

\begin{figure}[H]
	\centering
		\includegraphics[width=0.60\textwidth]{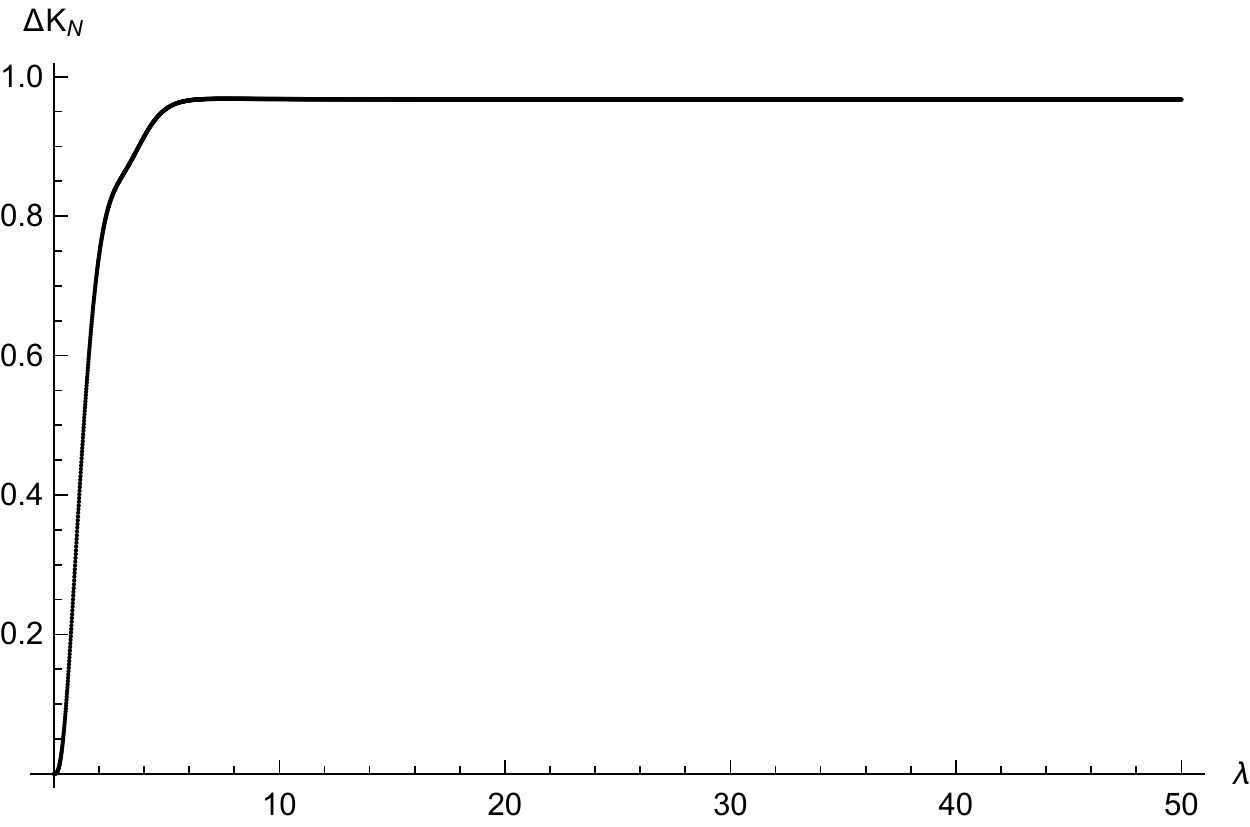}
	\caption{$\Delta K_N$ as a function of $\lambda$, considering $H_2$ and 
	     	initial conditions II.}
	\label{Figure-4}
\end{figure}

The behaviour above of $\Delta K_N$ is not always oscillating. In order to 
verify this feature, let us now consider $H_2$, with the initial conditions I
and II. The result is displayed in Figures (\ref{Figure-3}) and 
(\ref{Figure-4}).

\begin{figure}[H]
   \begin{minipage}{0.49\linewidth}
     \centering
     \includegraphics[width=\textwidth]{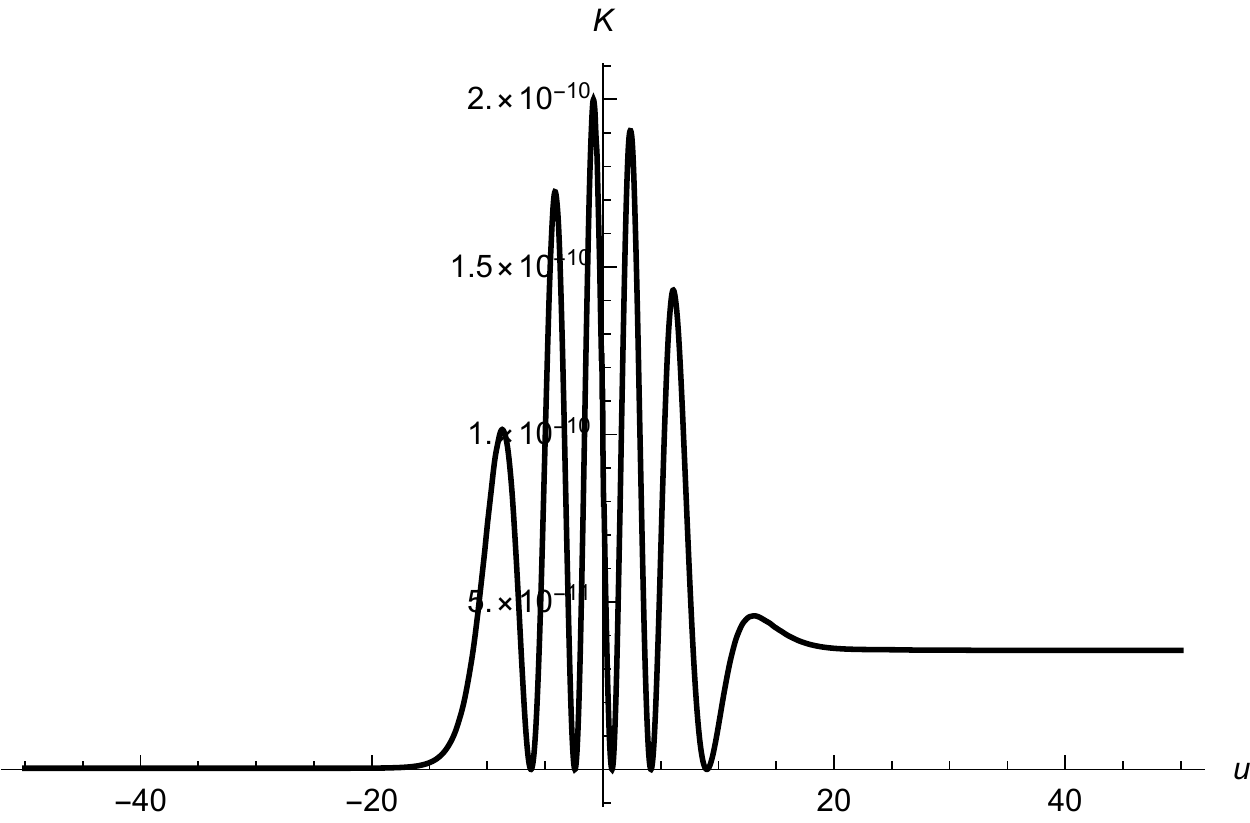}
     \caption{$K(u)$, considering $H_1$, initial conditions III and
          $\lambda =7.30247$.}\label{Figure-6}
   \end{minipage}\hfill
   \begin {minipage}{0.49\linewidth}
     \centering
     \includegraphics[width=\textwidth]{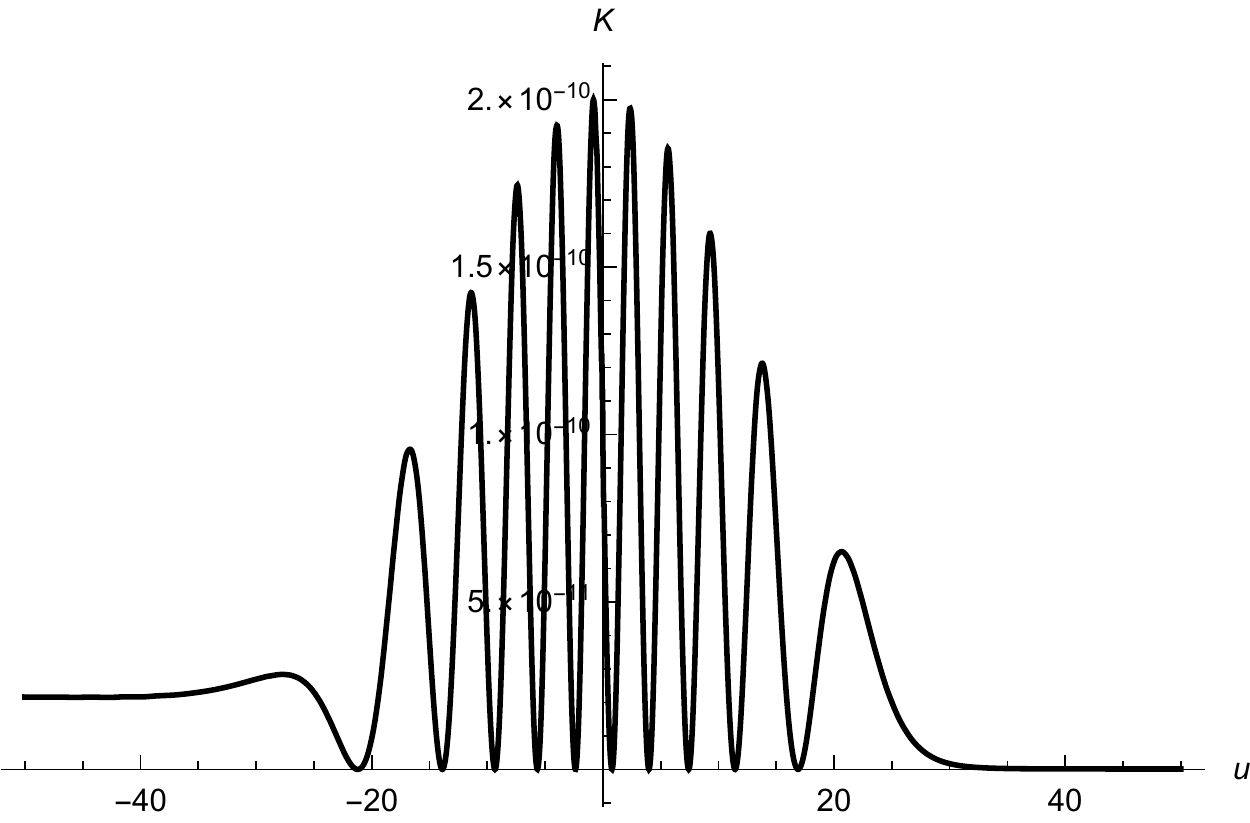}
     \caption{$K(u)$, considering $H_1$, initial conditions III and
          $\lambda =13.6043$.}\label{Figure-7}
   \end{minipage}
	\end{figure}

\begin{figure}[H]
   \begin{minipage}{0.49\linewidth}
     \centering
     \includegraphics[width=\textwidth]{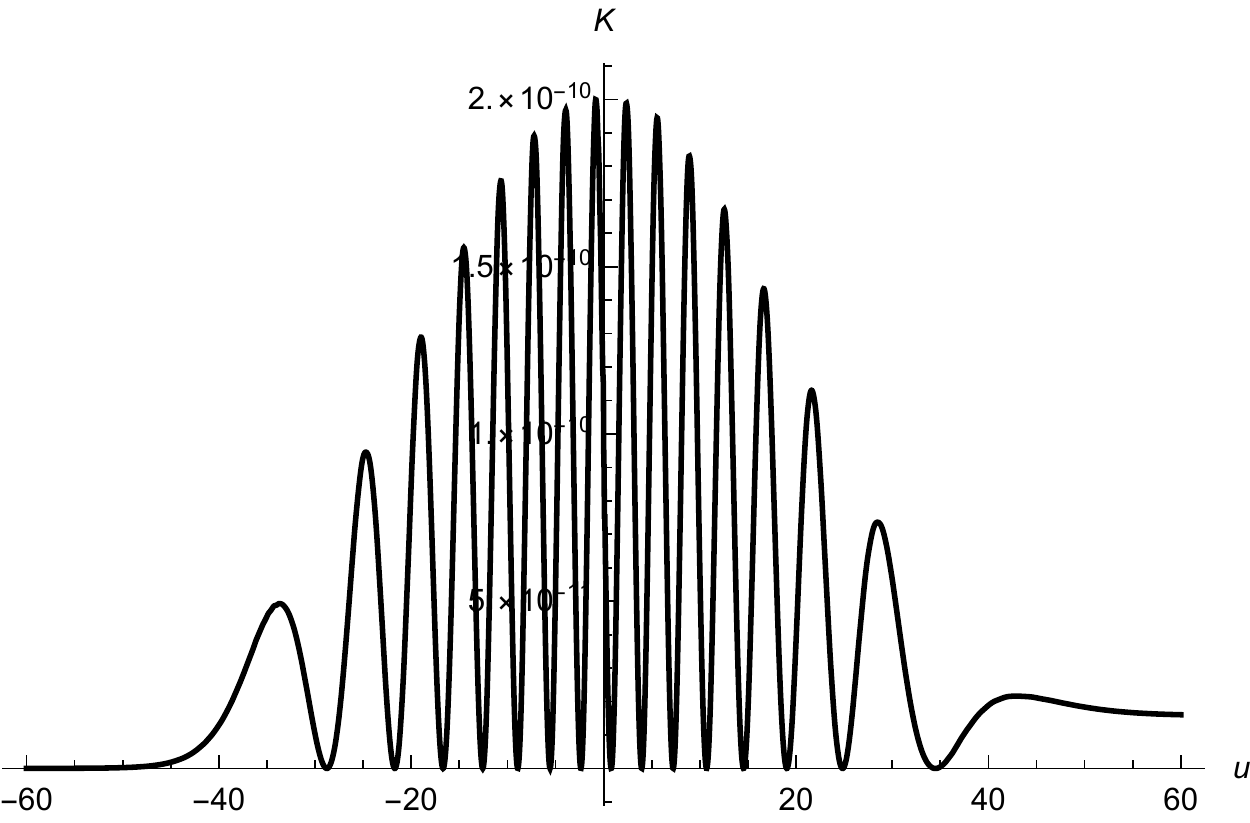}
     \caption{$K(u)$, considering $H_1$, initial conditions III and
          $\lambda =19.9062$. }\label{Figure-8}
   \end{minipage}\hfill
   \begin {minipage}{0.49\linewidth}
     \centering
     \includegraphics[width=\textwidth]{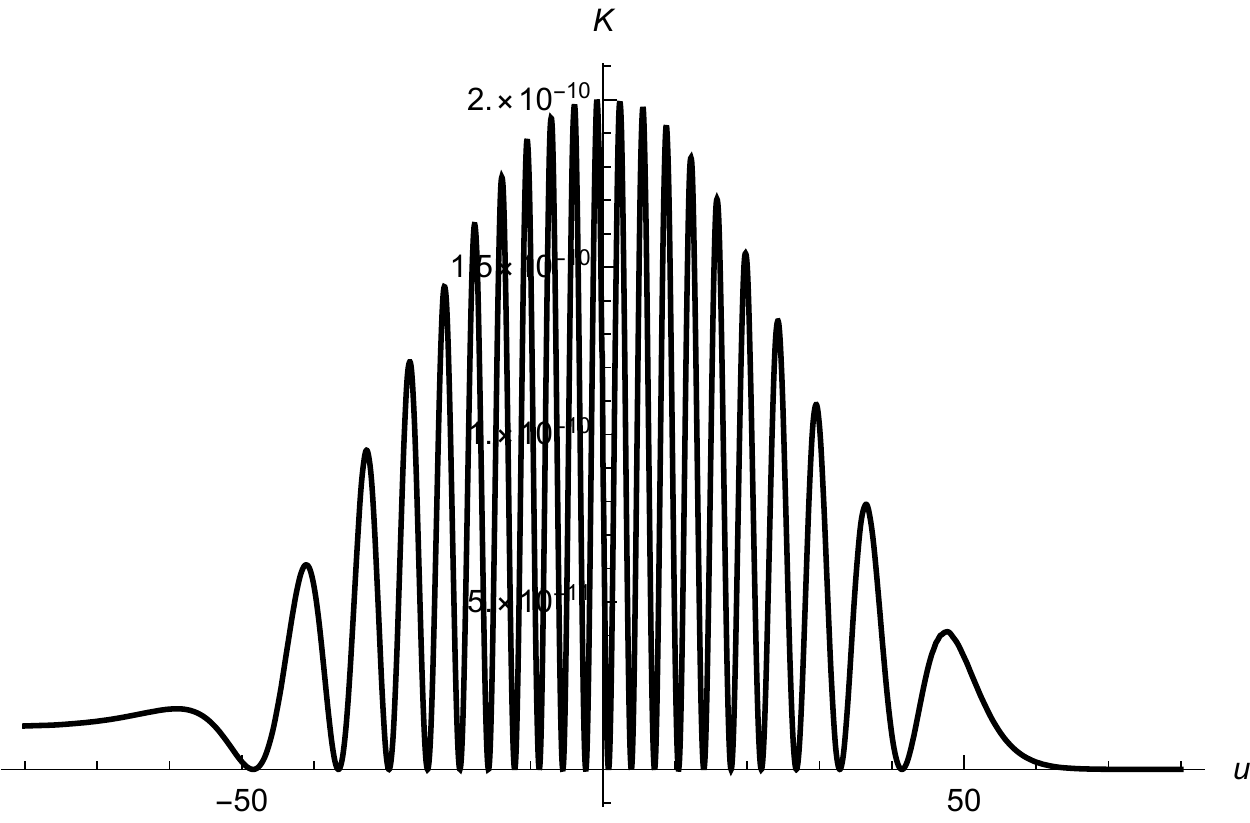}
     \caption{$K(u)$, considering $H_1$, initial conditions III and
          $\lambda =26.2081$.}\label{Figure-9}
   \end{minipage}
	\end{figure}

Next, we examine the behaviour of the kinetic energy $K$ as a function of $u$,
in order to probe its variation during the passage of the wave. For this 
purpose, we choose the function $H_1$ and the initial conditions III, 
considering 4 values of the width $\lambda$. The results are displayed in the 
figures (\ref{Figure-6})-(\ref{Figure-9}),

We see that the energy of the particle varies by a sequence of jumps and peaks, 
as if absorbing and releasing certain amounts of energy. One first attempt of 
explanation of these figures consists in assuming that the particle exchanges
small packages of energy with the gravitational field. In a hypothetical quantum
formulation of the problem, this would indeed be a natural assumption. The 
difference between any two peaks of energy would be a multiple of the quantum of
gravitational energy. Considering together Figures (\ref{Figure-1}), 
(\ref{Figure-2}), (\ref{Figure-6})-(\ref{Figure-9}), we may 
conclude that the particle is excited during the passage of the wave, around 
$u=0$, in a certain interval $\delta u$, and depending on the width $\lambda$ of 
the wave, the final state of the particle may be less energetic than the initial
state for reasons that are not yet clear, but we may consider the following
possibility: when the particle gains kinetic energy, the energy is extracted 
from the wave; when it looses kinetic energy, the particle produces 
gravitational radiation and increases the energy and intensity (although to a 
very small extent) of the incoming gravitational wave. In any case, 
we can certainly conclude that there is an actual exchange of energy between the
particle and the gravitational field. In the last section of the article we 
comment on the possible energy exchange between the gravitational and 
electromagnetic fields, according to investigations in the literature.

\subsection{Normalised gaussians}

Our main interest in this subsection is to verify the dependence of the physical
properties of the particle as we vary $\lambda$, when the particle is hit by a
normalised gravitational wave, and infer the results when $\lambda\rightarrow 0$
and the gaussian tends to a delta function. For this purpose, we will consider a
normalised wave determined by the function $H_3$,

\begin{equation}
H_3=-{1\over {10\lambda \sqrt{2\pi}}}
\rho^2(u)\lbrack 2\cos^2(\phi(u))-1\rbrack \,e^{-u^2/(2\lambda^2)}\,.
\label{25}
\end{equation}
All initial conditions in this subsection will be given by Eq. (\ref{19}), 
initial conditions I. We plot below, side by side, the kinetic energy of the
particle and the function $H_3$, both as a function of $u$, for 
$\lambda=0.05$, $\lambda=0.025$ and $\lambda= 0.01$. In the figure of the
function $H_3$, we make $\rho=\rho_0$ and $\phi=\phi_0$, given by initial 
conditions I.

\begin{figure}[H]
   \begin{minipage}{0.49\linewidth}
     \centering
     \includegraphics[width=\textwidth]{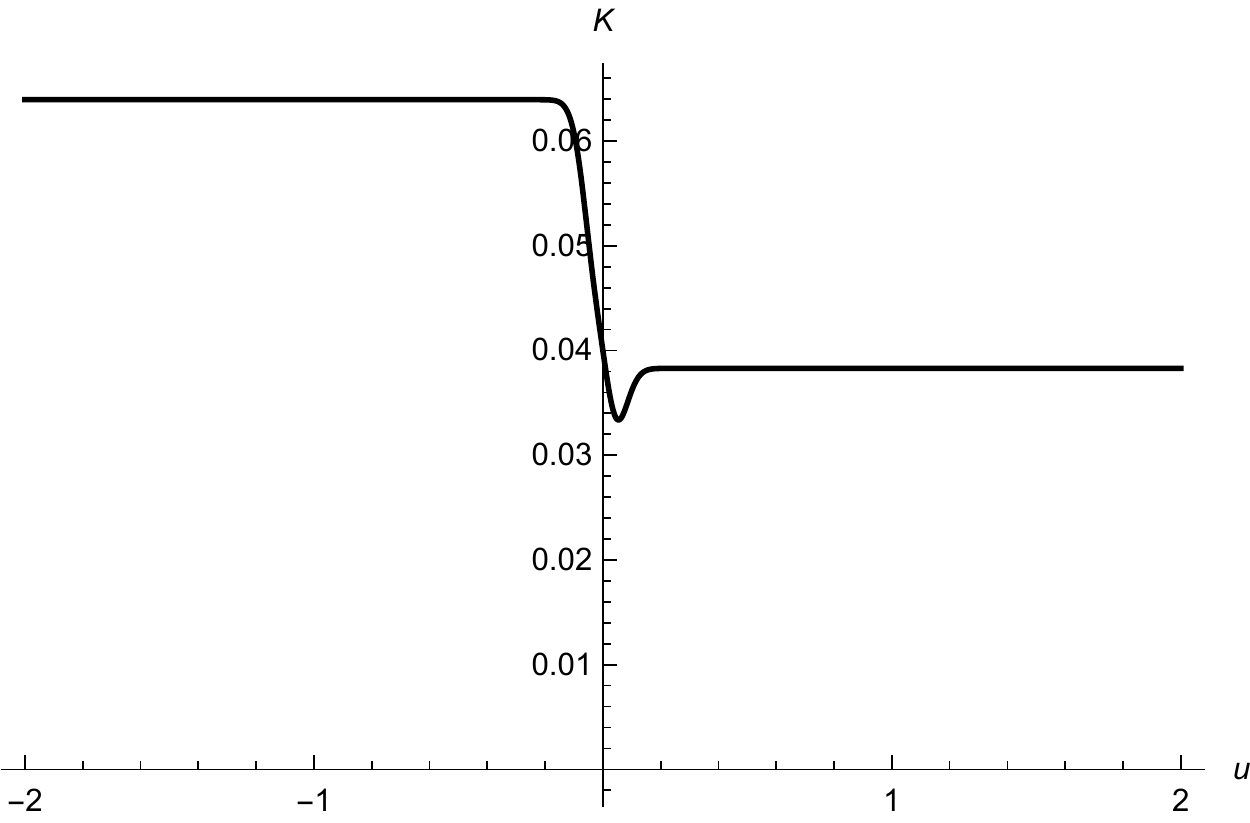}
     \caption{$K(u)$, considering $H_3$, initial conditions I and
          $\lambda =0.05$. }\label{Figure-10}
   \end{minipage}\hfill
   \begin {minipage}{0.49\linewidth}
     \centering
     \includegraphics[width=\textwidth]{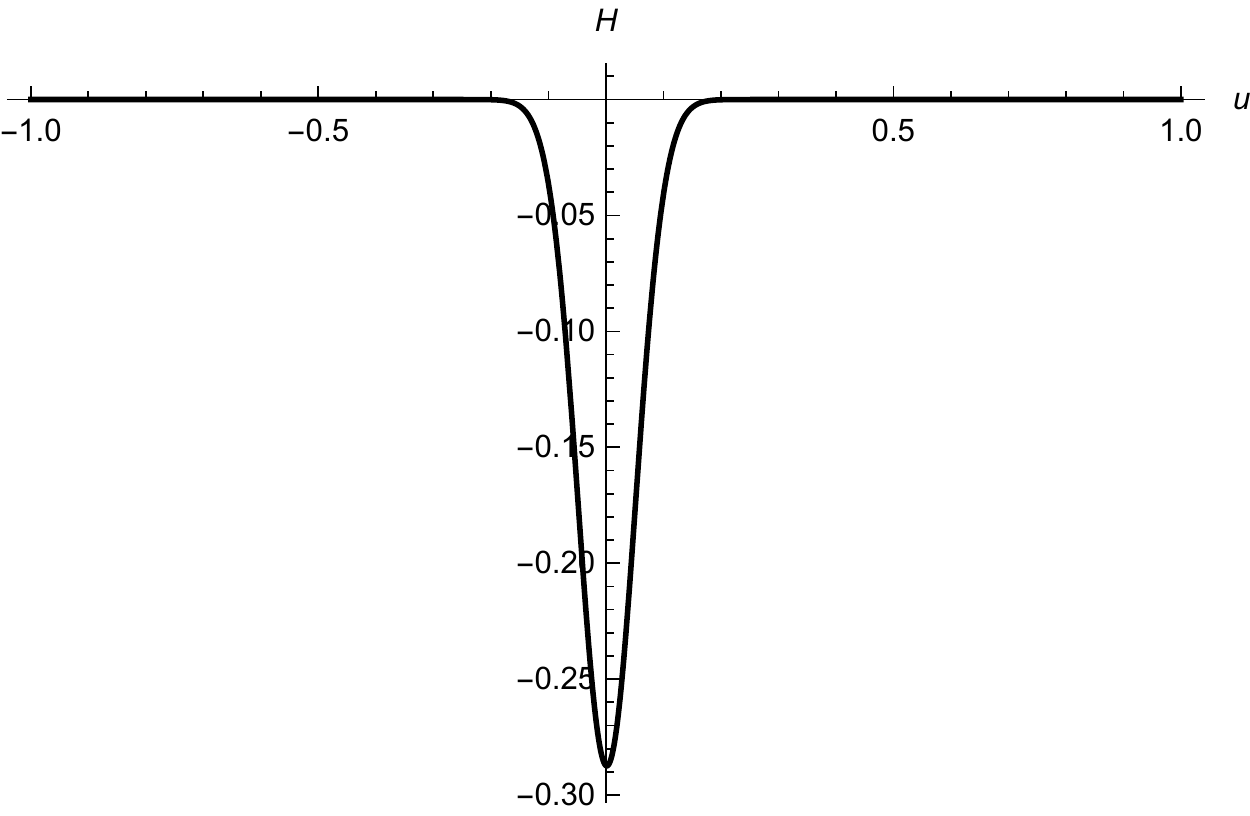}
     \caption{$H_3(u)$ when $\lambda =0.05$}\label{Figure-11}
   \end{minipage}
	\end{figure}

\begin{figure}[H]
   \begin{minipage}{0.49\linewidth}
     \centering
     \includegraphics[width=\textwidth]{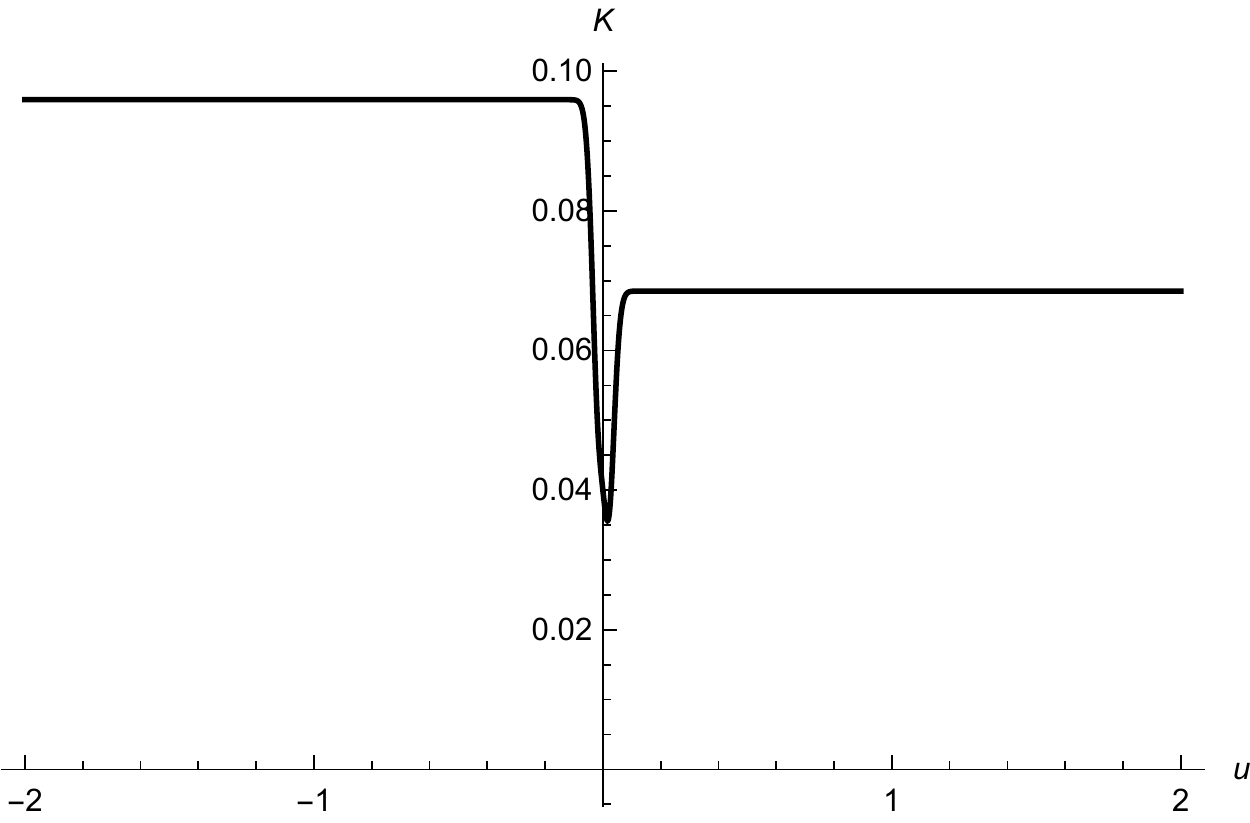}
     \caption{$K(u)$, considering $H_3$, initial conditions I and
          $\lambda =0.025$. }\label{Figure-12}
   \end{minipage}\hfill
   \begin {minipage}{0.49\linewidth}
     \centering
     \includegraphics[width=\textwidth]{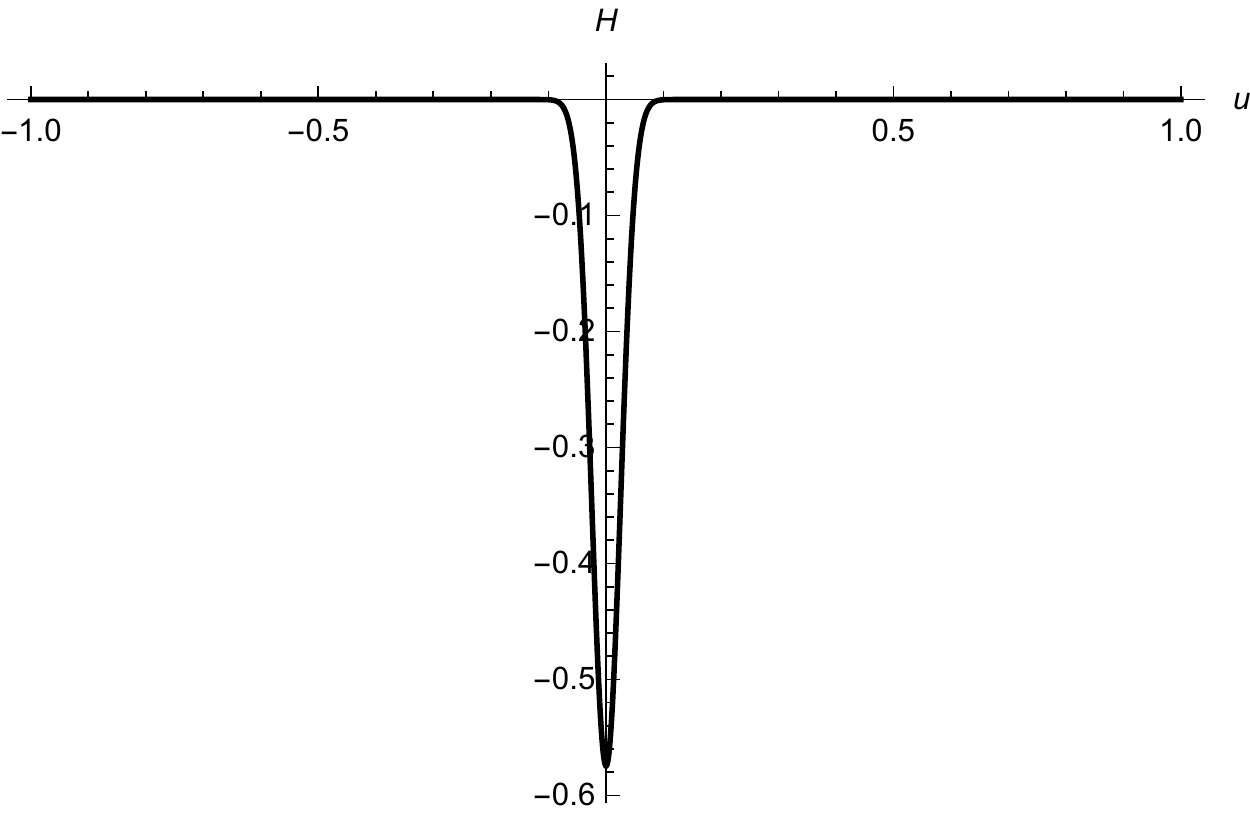}
     \caption{$H_3(u)$ when $\lambda =0.025$}\label{Figure-13}
   \end{minipage}
	\end{figure}

\begin{figure}[H]
   \begin{minipage}{0.49\linewidth}
     \centering
     \includegraphics[width=\textwidth]{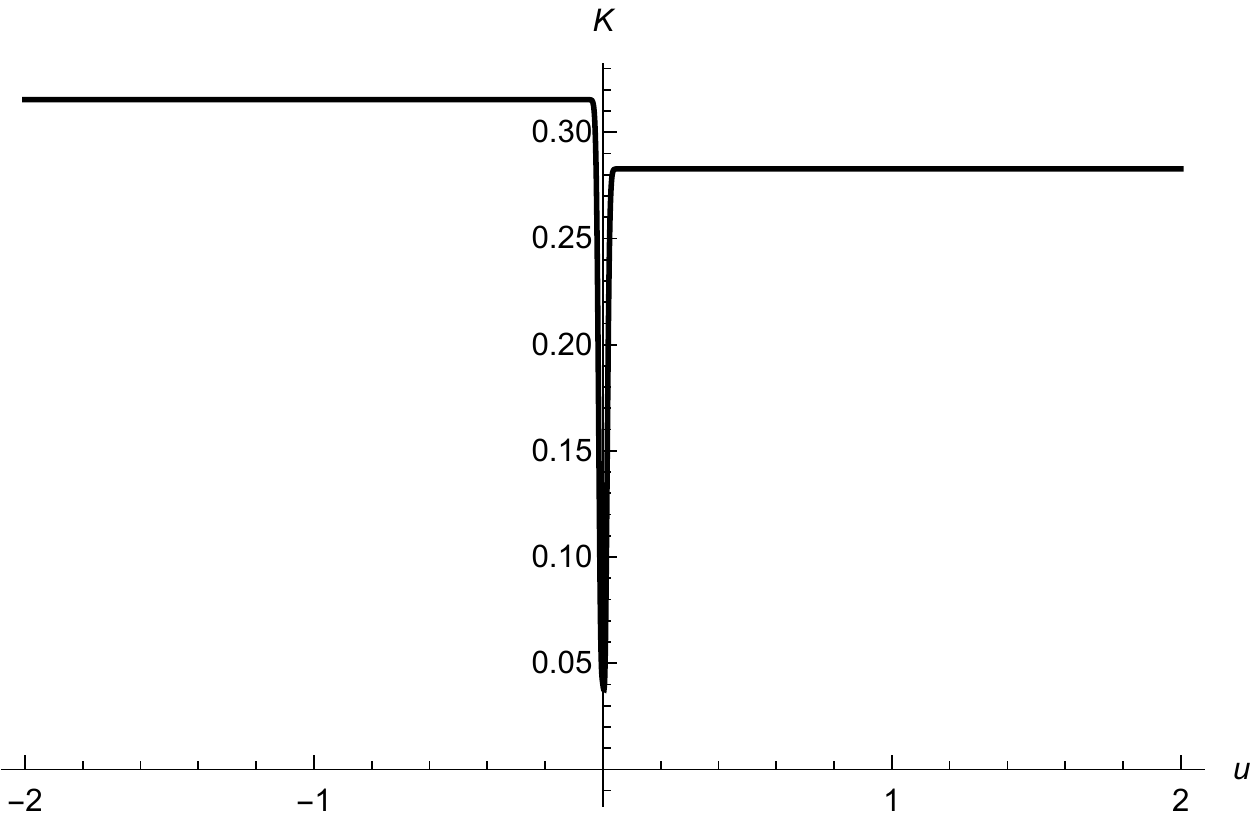}
     \caption{$K(u)$, considering $H_3$, initial conditions I and
          $\lambda =0.01$. }\label{Figure-14}
   \end{minipage}\hfill
   \begin {minipage}{0.49\linewidth}
     \centering
     \includegraphics[width=\textwidth]{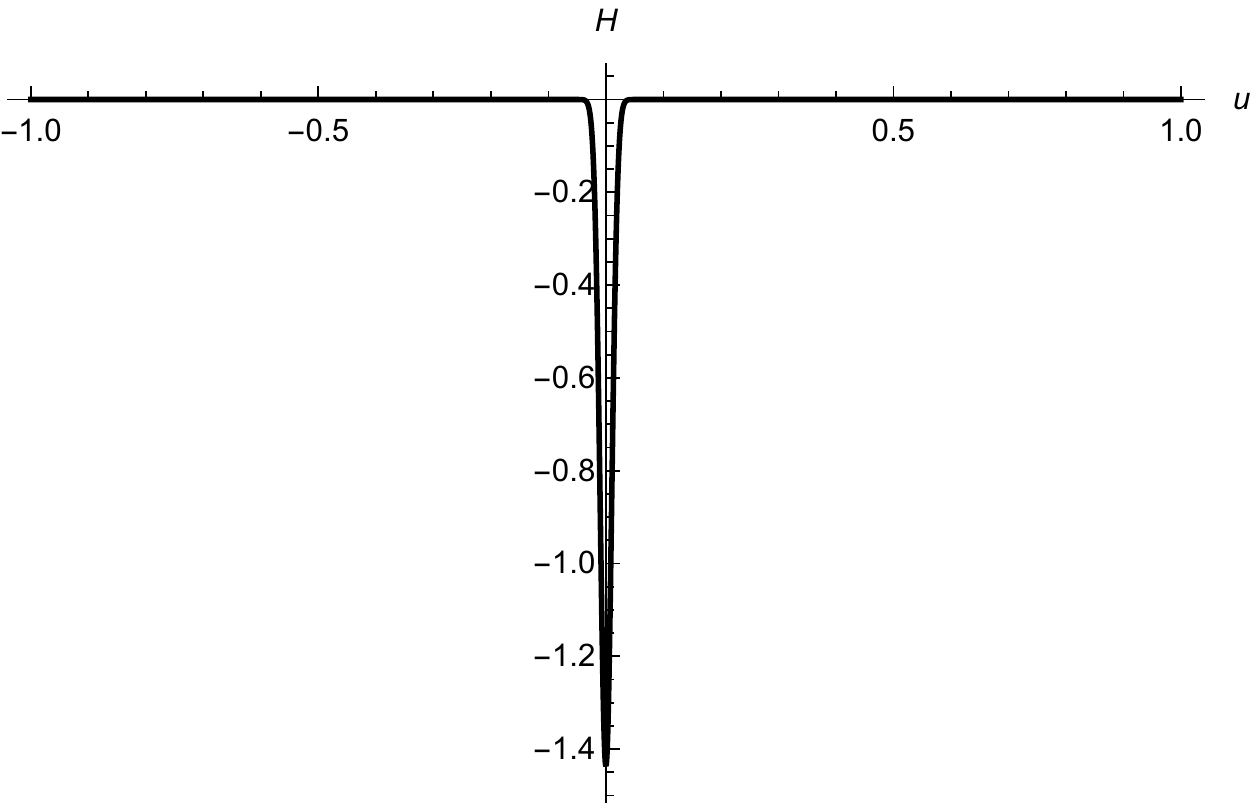}
     \caption{$H_3(u)$ when $\lambda =0.01$}\label{Figure-15}
   \end{minipage}
	\end{figure}

\begin{figure}[H]
	\centering
		\includegraphics[width=0.70\textwidth]{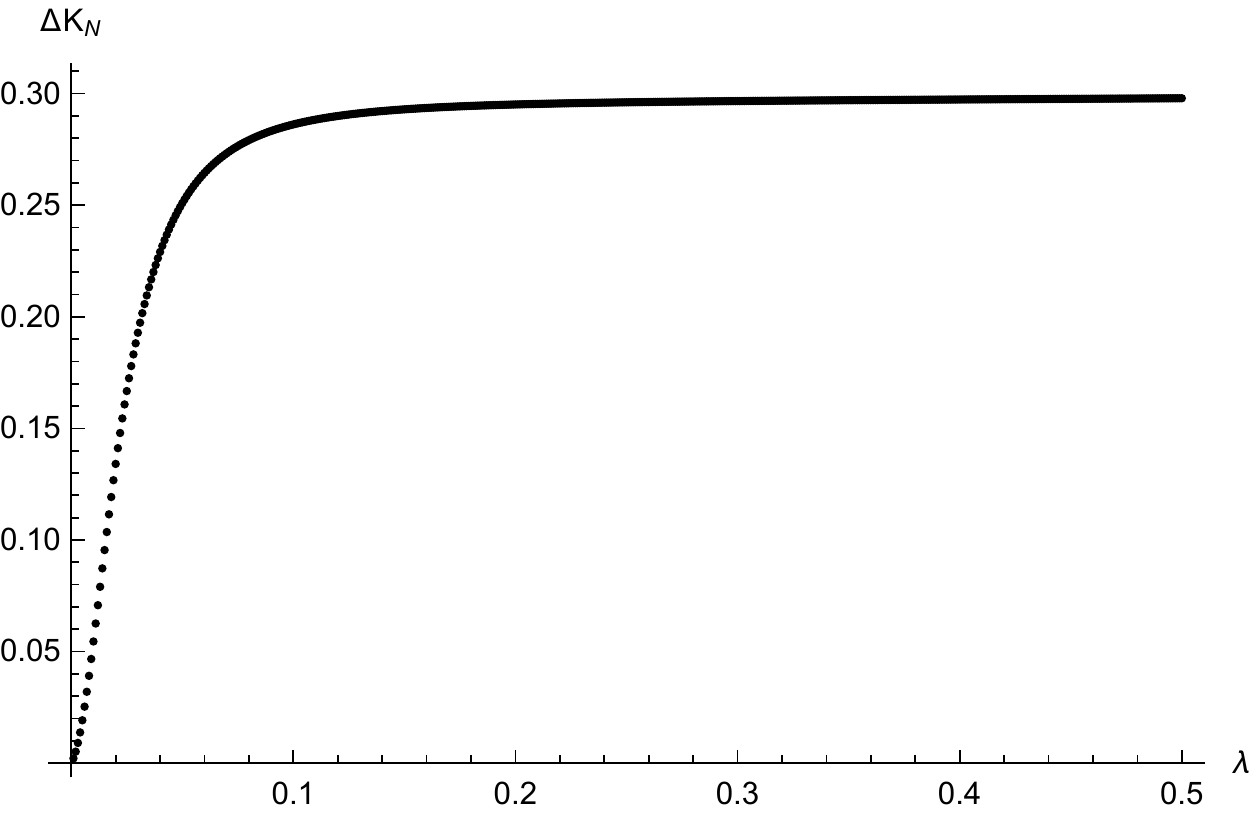}
	\caption{$\Delta K_N$ as a function of $\lambda$, considering $H_3$ and 
			initial conditions I.}
	\label{Figure-16}
\end{figure}

The sequence of Figures (\ref{Figure-11}), (\ref{Figure-13}) and 
(\ref{Figure-15}) clearly indicate that the functions $H_3(u)$ approaches a
(negative) delta function. 
By carefully observing the values of $K$ on the vertical axis of Figures 
(\ref{Figure-10}), (\ref{Figure-12}) and (\ref{Figure-14}), we see that the gap
$\Delta K$ between the final and initial energy of the particle increases as we
consider smaller and smaller values of $\lambda$. For the initial conditions 
considered, the absolute value of both $K_f$ and $K_i$ are increasing as we 
decrease $\lambda$. However, the normalised variation of the kinetic energy of 
the particle $\Delta K_N$ tends to zero when $\lambda \rightarrow 0$, according
to Figure (\ref{Figure-16}), and tends to a constant value for large values of
$\lambda$.

\begin{figure}[H]
	\centering
		\includegraphics[width=0.70\textwidth]{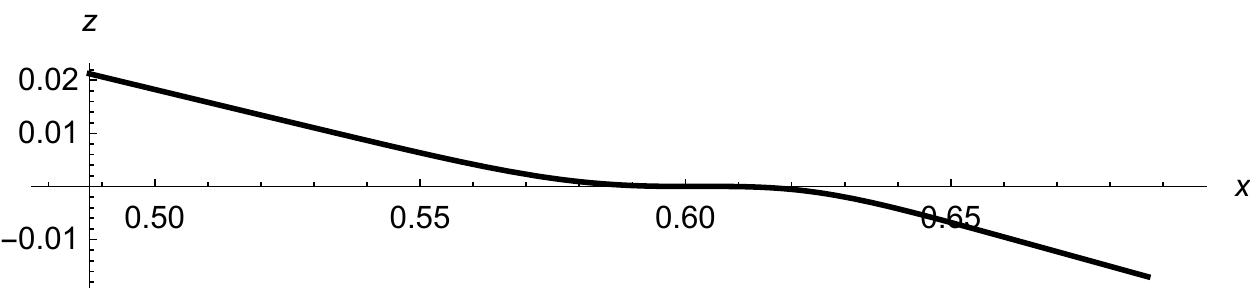}
	\caption{Trajectory of the particle in the $x-z$ plane, considering $H_3$, 
		initial conditions I, and $\lambda =0.1$.}
	\label{Figure-17}
\end{figure}

\begin{figure}[H]
	\centering
		\includegraphics[width=0.60\textwidth]{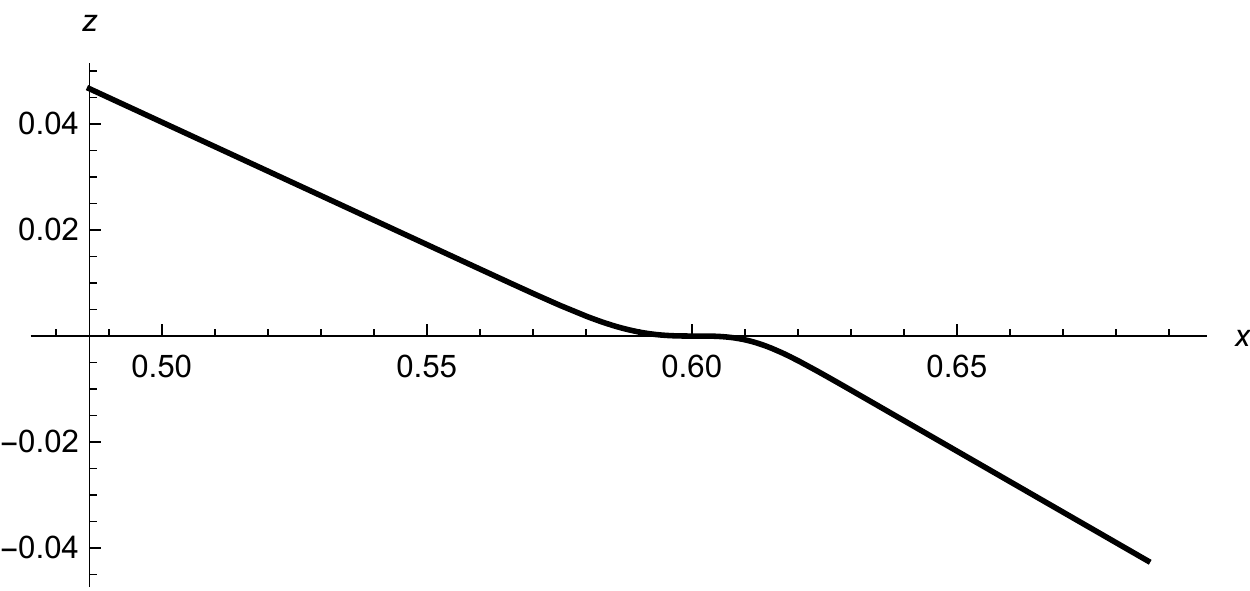}
	\caption{Trajectory of the particle in the $x-z$ plane, considering $H_3$,
		initial conditions I, and $\lambda =0.05$.}
	\label{Figure-18}
\end{figure}

\begin{figure}[H]
	\centering
		\includegraphics[width=0.60\textwidth]{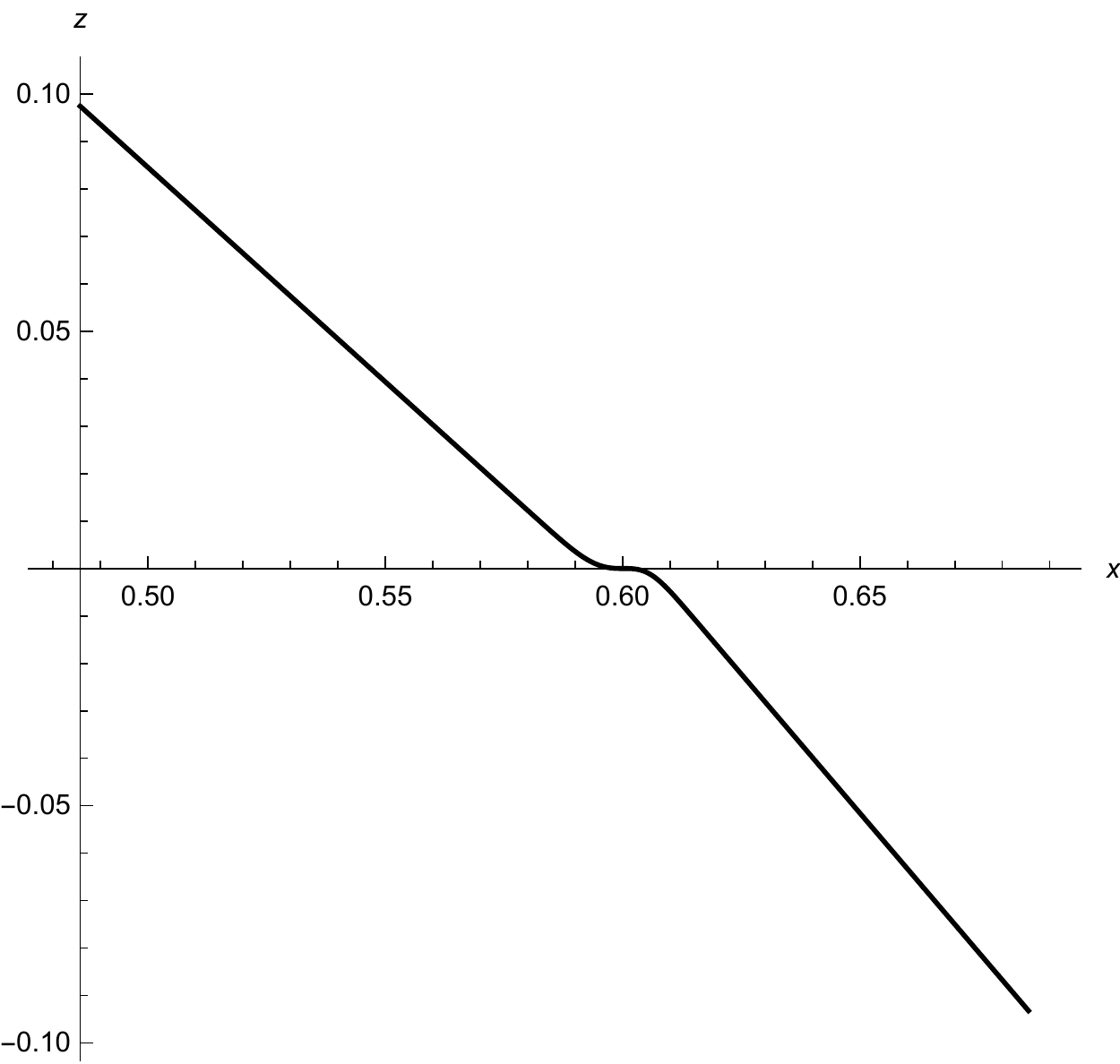}
	\caption{Trajectory of the particle in the $x-z$ plane, considering $H_3$,
	initial conditions I, and $\lambda = 0.025$.}
	\label{Figure-19}
\end{figure}

The feature discussed above, regarding the effect of the variation of $\lambda$
on $\Delta K_N$ also takes place in the geodesics of the particles. The particle
seems to be less sensitive to very small values of $\lambda$ (in fact, values of
$\lambda$ close to zero), as we can see in the 
Figures (\ref{Figure-17}), (\ref{Figure-18}) and (\ref{Figure-19}).

For simplicity of presentation, we have chosen to plot the trajectory of the 
particle only in the $x-z$ plane. We see that the smaller the value of
$\lambda$, the weaker is the effect of the gravitational wave on the trajectory
of the particle. Considering together Figures (\ref{Figure-2}), 
(\ref{Figure-3}), (\ref{Figure-4}) and (\ref{Figure-16}), we may conclude that 
$\Delta K_N \rightarrow 0$ when $\lambda \rightarrow 0$. This conclusion is 
likely to hold for normalised and non-normalised gaussians profile.

\section{Conclusions}

In this article we investigated particular features of free particles under the
action of pp-waves. We have paid special attention to the kinetic energy 
per unit mass of the
particles. The first conclusion is that there is an actual exchange of energy
(and momentum and angular momentum) between the particles and the gravitational
field of the wave. The second conclusion (that in fact was achieved in our 
previous investigations \cite{MRUC1,MRUC2}) is that the final energy of the 
particle may be smaller or higher than the initial energy, and that this feature
depends on the initial conditions of the free particles, that are not initially
at rest. The third conclusion is that the variation $\Delta K$ of the kinetic 
energy of the particles depends also on the width  $\lambda$ of the wave, and 
this dependence exhibits (not for all initial conditions) a surprising 
quasi-periodic behaviour. And finally, we have seen that 
$\Delta K_N \rightarrow 0$ when $\lambda \rightarrow 0$, in all cases 
investigated. This property should hold for all normalised gaussians. 
We assume that the variation of the kinetic energy of
the free particle is exactly minus the variation of the gravitational energy of
the wave. Therefore, the analysis of energy of free particles yields an indirect
evaluation of the variation of the gravitational energy.

We note that periodic
motion of free particles under the action of an exact, non-linear plane wave was
obtained in Ref. \cite{ZDGH3}, for suitable initial conditions of the particle.
Such periodic motion is obtained by directly introducing harmonic functions in 
the wave amplitude, and for this reason the resulting periodic trajectories in 
space-time bear no physical or mathematical 
relationship with the one described in Section 3.1. We also note
that recently an analytic investigation of $\Delta K = K_f - K_i$ has been
carried out in Ref. \cite{KASP}. The authors of the latter Reference adopt a
different function for the amplitude of the wave (which is not a 
gaussian), so that the geodesic equations may be solved analytically. They 
arrive at the same conclusion of Ref. \cite{MRUC1}, namely, that the final
kinetic energy of the free particle may be smaller or larger than the initial 
energy, after the passage of the pp-wave.

The higher is the value of $\lambda$, the longer (in time) is the interaction
of the particle with the wave, a feature that is suggested by 
Figures (\ref{Figure-6})-(\ref{Figure-9}). We establish the hypothesis that 
during the passage of the wave, the particle and the gravitational field 
exchange small packages of energy, and that the difference between any two peaks
of energy in Figures (\ref{Figure-6})-(\ref{Figure-9}) may be a multiple
of the quantum of gravitational energy, in a hypothetical quantum formulation.
Likewise, the energy gaps in Figures 
(\ref{Figure-10}), (\ref{Figure-12}) and (\ref{Figure-14}) should
also be a multiple of a minimum gravitational energy. The quantum of 
gravitational energy could be found, on speculative grounds, by means of a 
Millikan-type procedure (i.e., the procedure that allowed Millikan to 
determine the charge of the electron), after the analysis of a huge number of
figures. But maybe one would need to consider a quantised particle, instead of a
classical one. This issue will be investigated in the future, together with the
nature of the oscillations in Figures (\ref{Figure-1}) and (\ref{Figure-2}).

Throughout the article we
have analysed the energy exchange between gravitational waves and free
particles (i.e., particles that are subject only to the gravitational wave, but
which are otherwise free). However, the exchange of energy between gravitational
waves and electromagnetic fields has already been investigated in the 
literature, in the context of the linearised gravitational field.
(I) Gertsenshtein \cite{Gertsenshtein} showed that in the interaction of a 
gravitational wave with a static magnetic field, electromagnetic radiation may
be created. The energy of electromagnetic radiation is likely to come from the 
energy of the gravitational waves. (II) The graviton-photon interaction analysed
by Skobelev \cite{Skobelev} (graviton + graviton $\rightarrow$ photon + photon, 
photon + photon $\rightarrow$ graviton + graviton, 
photon + graviton $\rightarrow$ photon + graviton) also points out to an 
energy exchange between the gravitational and electromagnetic fields. (III) 
Similar investigations have been carried out by Jones {\it et. al.} 
\cite{Jones1,Jones2}, who studied particle production in a gravitational wave 
background and analysed the generation of electromagnetic radiation (photons) in 
the interaction of the gravitational wave with the quantum vacuum field 
fluctuations. The analyses in the four references above share similarities with the 
present investigation, in the sense that energy can be transferred between the 
gravitational waves (the quantum of gravitational energy) and the quantised 
electromagnetic field. The gravitational waves can gain or loose energy in these
processes. The localizability of the gravitational energy is a mandatory feature in
all these investigations. 

\bigskip

\noindent {\bf Acknowledgement}\par
\noindent We are grateful to an anonymous Reviewer for calling our attention to
Refs. \cite{Gertsenshtein}-\cite{Jones2}.

\end{document}